\documentclass[aps,prb,twocolumn,showpacs,floatfix]{revtex4}

\usepackage{float,graphicx}
\usepackage{psfig,verbatim}

\begin{document}

\title{Dynamic hysteresis from {\it zigzag} domain walls}

\author{Benedetta Cerruti$^{1}$ and Stefano Zapperi$^{1}$}

\affiliation{$^1$ CNR-INFM SMC and Dipartimento di Fisica, Universit\`a 
``La Sapienza'', P.le A. Moro 2, 00185 Roma, Italy}

\date{\today}

\begin{abstract}
We investigate dynamic hysteresis in ferromagnetic thin films with
{\it zigzag} domain walls. We introduce a discrete model describing 
the motion of a wall in a disordered ferromagnet  with in-plane 
magnetization, driven by an external magnetic field, considering 
the effects of dipolar interactions and anisotropy. 
We analyze the effects of external field frequency and temperature 
on the coercive field by Monte Carlo
simulations, and find a good agreement with the experimental data reported
in literature for Fe/GaAs films. This implies that dynamic hysteresis
in this case can be explained by a single propagating domain
wall model without invoking domain nucleation.
\end{abstract}
\pacs{77.80.Dj, 75.60.Ch, 75.70.Ak}
\maketitle

\section{introduction}

Ferromagnetic materials  are concrete 
examples of cooperatively interacting many-body systems. 
When a magnet is driven by a varying external magnetic field, 
the system cannot reach equilibrium instantaneously, due to the internal 
relaxation delay. If the external field oscillates, 
the magnetization will do so as well, lagging 
behind the field. This effect gives rise to a non-vanishing area
of magnetization-field loop whose form will depend on the 
applied field frequency: a phenomenon known as dynamic hysteresis 
\cite{chakrabarti,moore}.
Since the  loop area represents the amount of externally 
 supplied energy that is 
irreversibly transformed into heat during one magnetization cycle, 
dynamic hysteresis has important technological implications, e.g.
for high frequency devices 
applications. Furthermore, from a purely theoretical point of view,
the dynamics of disordered magnetic systems represents a central problem 
in non-equilibrium statistical mechanics. While in three dimensional systems 
dynamic hysteresis is well understood in terms of eddy currents dissipation,
this effect is expected to become negligible by reducing the sample 
thickness \cite{bertotti}. 
Thus recently there has been a renewed interest in understanding 
two dimensional systems, both experimentally 
\cite{moore,he,luse,jiang,raquet,suen,bland1,choi,lee2,lee3,moore2,chen,asti,santi,suen2}
and theoretically \cite{chakrabarti,zhong,zhong2,lyuksyutov}, 
motivated by the applications of thin ferromagnetic films in
magnetic recording technology and spintronic devices.

Two classes of models are mostly used to investigate the magnetization
reversal properties on a microscopic scale, spin models of the Ising
type \cite{chakrabarti,RFIM,rikvold1,rikvold2}, or extended domain wall models
\cite{zapperi,lyuksyutov,nattermann2001}.  The theoretical tools used to
interpret experimental data on dynamic hysteresis are often grown out
of the first class of models which suggest a universal scaling law 
for the dependence of the hysteresis loop area $A$
on the external parameters, i.e. the temperature of the
system $T$ and the amplitude $H_0$ and frequency $\omega$ of the
applied magnetic field. In particular, it is expected 
from the models that $A\propto
H_0^{\alpha}\omega^{\beta}T^{-\gamma}$, where $\alpha$, $\beta$ and
$\gamma$ are the scaling exponents \cite{chakrabarti}.  
The experimental estimates of these exponents display, however, a huge
variability
\cite{he,luse,jiang,raquet,suen,bland1,choi,lee2,lee3,suen2} and the
validity of that universal scaling law is still under debate \cite{santi,nistor}.  
Since various phenomena may in principle contribute to 
the hysteretic behavior, like domain
nucleation, domain wall propagation or simply retardation of
the magnetization due to fluctuations, it has been proposed that the
lack of good scaling of the function $A(\omega)$ is due to a
cross-over between two distinct dynamical regimes, one dominated by
domain wall propagation, and the other by nucleation of new domains
\cite{moore,bland1}.

The second class of models used to investigate ferromagnetic systems 
considers the dynamics of individual domain walls as the relevant
mechanism for hysteresis.  In two dimensional systems, 
developing such a kind of model can be 
even more complicated than in the bulk three dimensional case, 
due the possibility for the magnetization to lay in or out the film plane, and 
the huge variety of domains and domain walls topologies (for an exhaustive 
overview of the existing configurations together with many experimental 
images, see ref. \onlinecite{hubert}). Dynamic hysteresis due to the
motion of 180$^\circ$ Bloch domain walls has been extensively investigated
\cite{santi,lyuksyutov}, but less is know about charged walls.

In this article we focus on two dimensional systems with 
{\it zigzag} domain walls, arising from the competition between dipolar forces 
and magnetocrystalline anisotropy in thin films with head-on magnetization 
between nearest-neighbor domains \cite{freiser}. 
These walls have been originally observed  
in thin film magnetic recording media,
where head-on domains are induced by means of the application of a recording
head field, and have been then reported in several magnetic materials
such as iron \cite{iron}, Co \cite{co}, Gd-Co \cite{freiser}, 
epitaxial Fe films grown on GaAs(001) \cite{bland1}, {\it finemets} 
and many others. In addition, {\it zigzag} walls 
have also been observed  in ferroelectric materials, such as 
Gd$_2$(MoO$_4$)$_3$ crystals \cite{alexeyev}. 
Most calculations reported in literature for {\it zigzag} walls focus
on the derivation of the equilibrium parameters
(e.g. {\it zigzag} angle and amplitude or period) \cite{freiser,Eminima}
and do not consider their dynamics.

Here we introduce a discrete model for the motion 
of a single {\it zigzag} wall in a disordered ferromagnetic two
dimensional sample with in-plane uniaxial magnetization, 
driven by an external (triangular) magnetic field. The 
model is based on the interplay between dipolar and anisotropy 
energy contributions, in the presence of structural disorder. 
Dynamic hysteresis is investigated by Montecarlo simulations
analyzing the behavior of the coercive field $H_c$ as a function of
the external field frequency, temperature $T$, and sample thickness. 
We find good quantitative agreement with experimental data reported
for Fe/GaAs thin films \cite{bland1}.
Our results indicate that the experiments can be interpreted by a domain wall 
propagation model, and thus ruling out explanations involving a cross-over
with domain nucleation or other processes \cite{moore, bland1}.

The manuscript is organized as follows: in Sec. \ref{energetics} we present 
an overview on the energetics of a {\it zigzag} domain wall, computing 
magnetostatic, anisotropy and disorder energies. 
In Sec. \ref{comparison} we estimate the mean {\it zigzag} half-period and the coercive
field and compare our result with experimental observations, in order to test 
the reliability of our approximations. Next, in Sec. \ref{simulations} we present our 
model and the results obtained by Montecarlo simulations 
for the frequency, temperature and thickness dependence of the coercive field, and
compare them with experiments. Our results are finally resumed in Sec. \ref{conclusions}.

\section{energetics of {\it zigzag} domain walls}
\label{energetics}

In thin uniaxial ferromagnetic films, we can distinguish between 
two main classes of domain walls:
the first is represented by prevalently straight (magnetically uncharged) walls parallel 
to the easy axis, and the second by charged 
walls separating two domains with head-on magnetization. 
Since there is a cost of magnetostatic energy 
associated with the magnetic charge that increases with the sample
thickness, these walls are observed mostly in  
thin film \cite{hubert}. On the other hand, a charged straight wall 
is unstable and becomes metastable by forming a {\it zigzag} pattern 
to minimize its energy.

\begin{figure}[h]
\centerline{\psfig{figure=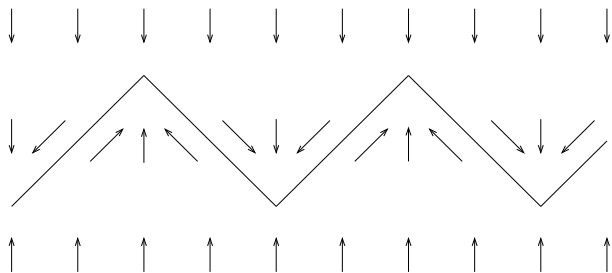,width=8.cm,clip=}}
\caption{The magnetic configuration in the N\'eel tail.}
\label{neel}
\end{figure}

A quite extensive derivation of the equilibrium {\it zigzag}
parameters (amplitude, period and angle) has been reported in
Ref.~\onlinecite{Eminima}. The calculation is based on a N\'eel tail
transition model \cite{sanders} which involves an in-plane
magnetization rotation over a transition region: the spins rotate
following the wall shape. Thus the entire region surrounding the wall
exhibits a nonuniform magnetization (see Fig.\ref{neel}).  This
spreading occurs at the cost of increasing the anisotropy energy.  The
total energy may be expressed as a function of the {\it zigzag}
parameters that are then obtained by minimization.  From our point of
view, it is important to stress that from magneto-optical images it can
be inferred that the zigzag angle is constant across the wall and does
not change during the motion.  Thus, in the following, we will consider
the angle as a fixed parameter in our model. As we mentioned in the
introduction, the {\it zigzag} shape of the wall is due to the
interplay between the magnetostatic and the anisotropy contributions
to the total energy \cite{freiser}.  The magnetostatic term opposes a
straight wall, which would maximize the magnetic charge density, and
favors a large {\it zigzag} amplitude, so that the magnetic charges
at the wall (all of the same sign) are as separated as possible.  The
anisotropy term prevents the amplitude to increase freely, avoiding an
excessive deviation of the magnetization from the easy axis
associated with a spread out N\'eel tail.

Our purpose is to study domain wall motion under an external magnetic field,
by discrete model simulations. To this end,  
we calculate the total energy of an arbitrary 
{\it zigzag} wall configuration.
As we are interested in the macroscopic response, we do not consider 
the details of the wall internal structure, and treat only
the magnetostatic, the anisotropy and the disorder contributions:
\begin{equation}
E=E_m+E_{an}+E_{dis}\mbox{.}
\label{E}
\end{equation}
In equation (\ref{E}), 
the magnetostatic term $E_m$ takes into account the interaction between 
magnetization and stray field, the anisotropy $E_{an}$ is the
energy cost of deviations from easy axis and $E_{dis}$ models
structural disorder, impurities, defects and so on. In the following 
subsections we will discuss these terms in more detail.

\subsection{Magnetostatic energy calculation}

\begin{figure}[h]
\centerline{\psfig{figure=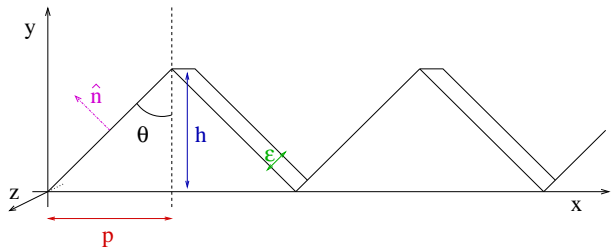,width=8.cm,clip=}}
\caption{(Color online) Sketch of the parameters of the {\it zigzag} wall. The easy axis is
along the $y$ direction.}
\label{sample}
\end{figure}

We consider two generic segments of a {\it zigzag} wall of total length $L$. 
We label the segments as $i=1,...,n$, where $n=L/p$ and $p$ is the 
half-period of the {\it zigzag}. We call $h$ the {\it zigzag} amplitude 
and $\theta$ the angle between the {\it zigzag} segment and the easy axis.
The thickness of the film is $\epsilon$, and it coincides with the wall 
thickness as we will consider only rigid walls 
(see Fig. \ref{sample} for a definition of the parameters). 

Since the magnetostatic self-energy only depends on the total magnetic 
charge, which is constant during the wall motion, we limit the 
calculation to the magnetostatic interaction energy (i.e. $i\neq j$).
The contribution due to the $i$-th and the $j$-th segments 
may be written as the surface integral
$$E_{ij}=\int\sigma\phi_i(x,y,z)\mbox{d}S_j\,,$$
where $\sigma={\bf M}\cdot\hat{{\bf n}}=2M_s \sin \theta$ is the (constant) 
surface magnetic charge density ($M_s$ is the saturation 
magnetization), $\hat{{\bf n}}$ is the unit vector normal to wall segment
surface and $S_j$ is the surface of the $j$-th segment. 
The scalar potential generated by the $i$-th segment, $\phi_i(x,y,z)$,
 is calculated by 
$$\phi_i(x,y,z)=\int\frac{\sigma}{|{\bf r}-{\bf r}'|}\mbox{d}S_i'.$$ 
So, for $\epsilon$ small with respect to the segments distance, we can write
\begin{eqnarray}
E_{ij} &=& 8M_s^2\epsilon^2\mu_0\int_{jp}^{(j+1)p}\mbox{d}x\int_{ip}^{(i+1)p}\mbox{d}x'
\nonumber\\
&&\frac{1}{\sqrt{(x-x')^2+(m_ix+q_i-m_jx'-q_j)^2}},
\nonumber\\
\label{Eij}
\end{eqnarray}
where $m_i$ and $m_j$ are the slopes (which values should be 
$\pm h/p=\pm\tan(\pi/2-\theta)\,$) 
and $q_i$ and $q_j$ the $y$-intercepts 
of the $i$-th and $j$-th segments. 
The direct solution of equation (\ref{Eij}) 
is very involved, and we report it in the appendix.

\subsection{Anisotropy energy calculation}

The anisotropy energy term $E_{an}$ 
describes the energy cost of the deviation of 
magnetic moments from the easy axis of the material, which 
in the simple case of an uniaxial crystal can be written as
\begin{equation}
E_{an}=\int\!\!\! d^3r K_u \sin^2\phi  ,
\label{anisotropy}
\end{equation}
where $K_u$ is the in-plane 
uniaxial anisotropy constant and $\phi$ is the angle 
between the easy axis and the magnetization vector.
The rotation of the magnetization is associated with the N\'eel
tail (Fig.\ref{neel}). We assume \cite{freiser} that the charge is 
uniformly distributed over the entire band containing the wall. Although this 
approximation exaggerates the diffuseness of the charge, it has been used
to calculate vertex angles and {\it zigzag} amplitudes that resulted 
to be in reasonable agreement with experimental observations \cite{freiser}.
We can estimate Eq. \ref{anisotropy} (with the notation sketched in Fig. 
\ref{sample}) and obtain:
\begin{equation}
E_{an}=\frac{\epsilon K_u}{h \tan{\theta}}\int_{-h/2}^{h/2} dy \int_{-(h/2-y)\tan{\theta}}^{(h/2-y)\tan{\theta}} dx \sin^2\phi(x,y) {.}
\label{phi}
\end{equation}

Assuming that $\phi(x,y)$ describes a linear in-plane rotation of the 
magnetization vector, $\phi(x,y)=\theta x/\left[(h/2-y)\tan\theta\right]$, 
developing $\sin^2\phi$ in power series and then integrating term by term, 
we obtain the anisotropy energy for unit 
length as
\begin{equation}
E_{an}=\epsilon K_u h c(\theta) ,
\label{E_an}
\end{equation}
where $c(\theta)$ is a constant function of the {\it zigzag} angle $\theta$:
$$c(\theta)=\sum_{m,l=0}^{\infty}\frac{(-1)^{m,l}}{(2m+1)!(2l+1)!}\,
\frac{\theta^{2(m+l+1)}}{2(m+l+1)+1} {,}$$
which could be evaluated numerically.

\subsection{Disorder}

Different sources of inhomogeneities are found in virtually all
ferromagnetic materials, and the presence of structural disorder is
essential to understand the hysteretic behavior, especially to account
for the residual coercive field at $\omega\rightarrow 0$. Disorder is
provided by vacancies, non-magnetic impurities, dislocations or grain
boundaries in crystalline systems, variations of the easy axis between
different grains for polycrystalline materials and internal stresses
for amorphous alloys. We will consider only quenched (frozen)
disorder, that does not evolve on the timescale of the magnetization
reversal.  For simplicity we model disorder by an energy contribution 
associated to each site which may be occupied by a segment
(our discrete unit length) of the {\it zigzag} wall. This
term is extracted from an uncorrelated random Gaussian distribution with zero
mean.

The dependence of the disorder energy  on the film thickness 
can be obtained by simple considerations. Given that the
minimal segment has length $a$, we discretize the sample thickness  
in $\epsilon/a$ elements and associate a Gaussian random variable to
each of them. The mean square value of the disorder energy 
per unit domain wall length is then proportional to the sum of contribution 
over the whole thickness, and thus
to $\epsilon$. As a consequence of this, the disorder contributes 
to the total energy  
(eq. (\ref{E})) by a term proportional to the square root of the 
film thickness
\begin{equation}
E_{dis}= \sqrt{\epsilon}\Delta,
\label{Edis}
\end{equation}
where $\Delta$ is the variance of the random variables.

\section{Theoretical considerations and comparison with materials}
\label{comparison}

\subsection{{\it Zigzag} parameters}

A way to test the reliability of our model is to compare 
the results of our model for some relevant parameters with the measured
experimental values. As an example, we can estimate the typical {\it zigzag}
half-period  $p_{eq}$ for an equilibrium configuration  
in absence of external field. An  approximation of the magnetostatic energy can be obtained
in closed form by developing Eq. \ref{Eij} for $p<\!\!<L$ and is given, for unit length, by
$$E_m\simeq 8\mu_0 M_s^2\epsilon^2\ln{(L/p)}$$ 
In the absence of disorder and  for $H_{ext}=0$, the total energy (Eq.~\ref{E}) can be written as
$$E=E_{m}+E_{an}\simeq 8M_s^2\epsilon^2\mu_0\ln{(L/p)}+\epsilon K_u\frac{p}{\tan{\theta}}c(\theta),$$
where we have imposed $h=p/\tan{\theta}$. If the configuration is stable, considering 
the $T=0$ case, we can impose $\partial E/\partial p=0$ and obtain 
$$-\frac{8M_s^2\epsilon^2\mu_o}{p}+\frac{\epsilon K_uc(\theta)}{\tan{\theta}}=0,$$
from which follows
\begin{equation}
p_{eq}=\frac{8M_s^2\epsilon\mu_0\tan{\theta}}{K_uc(\theta)}.
\label{p_eq}
\end{equation}

We can estimate the numerical value of $p_{eq}$ by using the parameters 
reported in literature for typical ferromagnetic thin films.
For example, for $Fe/GaAs(001)$ analyzed in Ref. \onlinecite{bland1} we can 
set $\mu_0M_s=2T$ and $K_u=0.5\times 10^5 J/m^3$, 
so that for a thickness $\epsilon=25 nm$ and an angle $\theta=20^\circ$, we 
obtain
$$p_{eq}\simeq 100 \,\,\mu m,$$
which is in good agreement with the typical lengthscale inferred
from magneto-optical investigations \cite{bland1}.

\subsection{Coercive field and thickness dependence}
\label{epsi}

Another quantity that is interesting to obtain quantitatively is the 
value of the (zero frequency) coercive field. A very rough order of magnitude 
estimate could be obtained supposing that the disorder is small and that the 
anisotropy and magnetostatic  terms are of the same order of magnitude. 
Close to the coercive field when the 
energy variation is stationary we can set
$$\mu_0M_s\epsilon phH_c=\epsilon p h K_uc(\theta),$$
which implies
\begin{equation}
H_c=\frac{K_uc(\theta)}{\mu_0M_s}.
\label{eq:hc_k}
\end{equation}
Employing the parameter values reported
above, we obtain $H_c\sim 15 \,\,Oe$, which is in reasonable 
agreement with the range of  values found for Fe/GaAs thin films \cite{bland1}.

A more refined quantitative calculation of $H_c$ is difficult because we can not easily 
quantify the value of the disorder term. Nevertheless, we can obtain from simple
considerations the thickness dependence of coercive field.
The energy difference between two {\it zigzag} wall 
configurations in the presence of external magnetic field $H_{ext}$ is given by (see eq. \ref{E}):
\begin{equation}
\Delta E=\Delta E_m+\Delta E_{an}+\Delta E_{dis}\mbox+\Delta E_{ext}{,}
\label{deltaE}
\end{equation}
where the interaction $\Delta E_{ext}$ with $H_{ext}$ is given by
\begin{equation}
\Delta E_{ext}=-2\mu_0 M_s\epsilon H_{ext}\Delta A{,}
\label{Eext}
\end{equation}
and $\Delta A$ is the area interested by the magnetization reversal.
As it could be seen from equations \ref{Eij}, \ref{E_an} and 
\ref{Edis}, eq. \ref{deltaE} could be rewritten as
\begin{equation}
\Delta E=\epsilon^2 \Delta E_m'+\epsilon \Delta E_{an}'+\sqrt{\epsilon} \Delta E_{dis}'+\epsilon\Delta E_{ext}',
\label{eps_riscalati}
\end{equation}
where $\Delta E_m'=\Delta E_m/\epsilon^2$, 
$\Delta E_{an}'=\Delta E_{an}/\epsilon$, 
$\Delta E_{dis}'=\Delta E_{dis}/\sqrt{\epsilon}$
and $\Delta E_{ext}'=\Delta E_{ext}/\epsilon$ do not dependent on $\epsilon$.
As $\Delta E$ represents the energy barrier 
that the {\it zigzag} wall has to overcome in order to move in the direction of the external 
magnetic field, we expect that Eq.~\ref{eps_riscalati} encodes
crucial informations on the dynamics. 

The coercive field dependence on the thickness, can be obtained comparing
$\Delta E_m$ with the most relevant of the various terms contributing to $\Delta E$. 
For $\epsilon \rightarrow 0$, $\Delta E$ is dominated by the 
disorder contribution. Comparing $\Delta E_m \propto \epsilon^2$ with 
$\Delta E_{dis} \propto \sqrt{\epsilon}$, we easily obtain $H_c \propto 1/\sqrt{\epsilon}$.
If the wall is strongly pinned  by disorder or anisotropy, starting from a 
saturated configuration, small external field 
changes are not able to trigger avalanches, resulting in square-shaped hysteresis loops 
with high coercitivity. Otherwise, if the pinning is weak, avalanches are induced even by relatively
small field changes, so that the loops will be tighter, with small coercitivity. We can expect that the 
$1/\sqrt{\epsilon}$ dependence will disappear in the limit of weak disorder regime.
For intermediate $\epsilon$ values, the leading terms in eq. \ref{eps_riscalati} will be the 
anisotropy energy $\Delta E_{an} \propto \epsilon$ and the interaction with 
the external magnetic field $\Delta E_{ext}$, leading to a thickness independent
coercive field given by Eq.~\ref{eq:hc_k}.
Finally, for larger values of $\epsilon$, the dipolar energy will lead the 
energy barrier balance, inducing a linear dependence 
thickness dependence of the coercive field ($H_c \propto \epsilon$).
As it is shown in the following  sections, these general considerations are confirmed by 
numerical simulations.

\section{montecarlo simulations}
\label{simulations}

\subsection{The model}

\begin{figure}[h]
\centerline{\psfig{figure=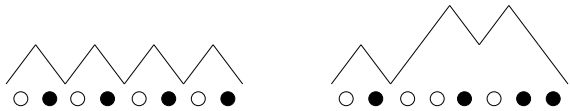,width=7.5cm,clip=}}
\caption{The mapping. Two examples of {\it zigzag} configuration.}
\label{zigzag}
\end{figure}

To investigate dynamic hysteresis, we perform Montecarlo simulations based on 
the energy terms derived above. 
We consider a wall of length $L$ in a sample of finite dimensions. 
Since we are interested in macroscopic effects, 
we discretize the wall defining a minimal segment length 
$a$ and  map the {\it zigzag} wall into a particle model.
We define the minimal elements with negative slope as a  particle 
and those with positive slope as a void, as sketched in Fig.\ref{zigzag}. 
Thus we reduce the two-dimensional problem of a {\it zigzag} wall into
a one-dimensional particle model evolving under the appropriate
dynamic rules.

The Montecarlo dynamics is implemented by choosing 
randomly an active pair of nearest neighbor elements, i.e. a particle-void or a void-particle pair,
and trying to exchange their positions between each other.
This rule corresponds to allow only the motion 
of segments with down-up or up-down slope, preserving 
the zigzag (solid-on-solid) structure of the wall.
Once a possible displacement has been attempted, we calculate the energy 
difference  (see Eq. \ref{deltaE}) between  
the starting configuration and the new one. 
If $\Delta E\le 0$ we accept directly the move, otherwise the move is
accepted with probability $P=\exp[-\Delta E/k_B T]$.

In practice, the various contributions to $\Delta E$ are evaluated as follows:
\begin{itemize}
\item $\Delta E_m$ is obtained from Eq. (\ref{Eij}). To simplify the
expression, we can use $E_{ij}=1/r_{ij}$ if the particles 
are not nearest neighbor, and  the whole expression in Eq. (\ref{risultato}) 
otherwise. As it is discussed in the appendix, 
$E_{ij}$ deviates significantly from $1/r_{ij}$ only 
if $i$ and $j$ are nearest-neighbor. We can then perform another simplification, by absorbing	
the deviation from $1/r_{ij}$ into the anisotropy (nearest-neighbor) term.
In summary we can set $E_{ij}$ equal to $8 \mu_0 M_s^2 \epsilon^2/r_{ij}$ $\forall i,j$,
and renormalize the anisotropy term by an appropriate magnetostatic constant. Notice that
$\Delta E_m$ is an attractive long range term, tending to aggregate all
the particles. This configuration would correspond to a single period of a 
{\it zigzag} with $p=L/2$. 
\item $\Delta E_{an}$ is a nearest-neighbor repulsive term
that favors configurations where particles are followed by voids.
This prevents the formation of a {\it zigzag} with a wide 
amplitude. In practice this term is treated as a positive contribution
if the left (right) nearest neighbor of the segment $i$ ($i+1$)
to be flipped has opposite slope with
respect to $i$ ($i+1$), and a vanishing contribution otherwise. In this way 
we treat the rotation of the magnetic moments close to the 
{\it zigzag} wall, as uniformly distributed in a ``band'' surrounding the 
wall \cite{freiser}. 
\item $\Delta E_{dis}$ represents the contribution from structural 
disorder, whose time-independent 
intensity is randomly extracted with Gaussian distribution with zero mean for every site.
\item $\Delta E_{ext}$, the interaction with the external magnetic field $H_{ext}$, is calculated from Eq.~\ref{Eext},
i.e. $\Delta E_{ext}=-2\mu_0 H_{ext}\Delta M$, where $\Delta M$ is the 
magnetization difference between the two configurations.
\end{itemize}

If the move is accepted we update the configuration and continue the
process for a time interval $\Delta t$. In the spirit of the Montecarlo
method, each attempt corresponds to a time step $\tau =1/N_{act}$, where
$N_{act}$ is the number of active particles (i.e. those that are followed
or preceded by a void). After each time interval $\Delta t$ is expired we increase the
external field by $\Delta H_{ext}$  and restart the updating process.
This rule corresponds to a field rate 
$\dot{H}\equiv\Delta H_{ext}/\Delta t \propto \omega$.
We begin the simulation from the 
$M=0$ at $H_{ext}=0$ state and drive the sample to 
positive and then to negative saturation.  

\subsection{Results}

\subsubsection{Frequency dependence}

\begin{figure}[h]
\centerline{\psfig{figure=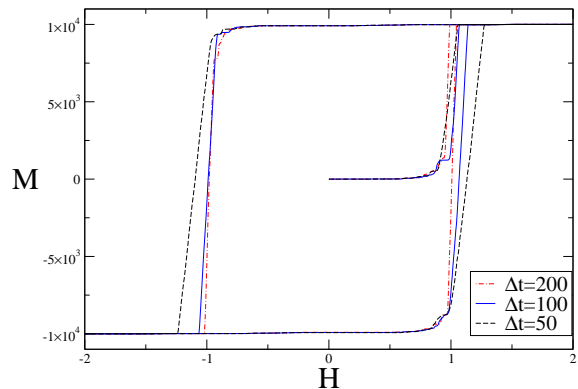,width=8.4cm,angle=-90,clip=}}
\caption{(Color online) Some $T=0$ hysteresis cycles for different external field time rate.}
\label{T=0}
\end{figure}

\begin{figure}[h]
\centerline{\psfig{figure=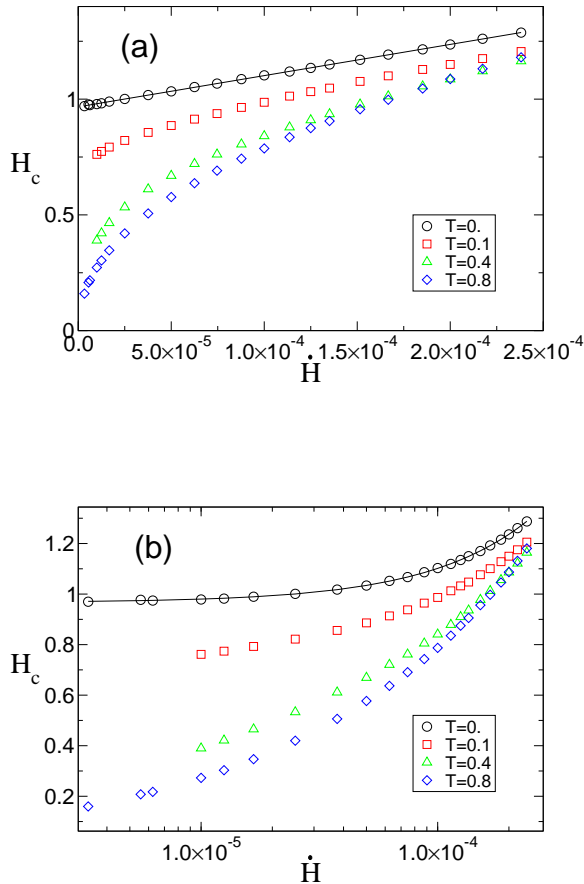,width=8cm,angle=-90,clip=}}
\caption{(Color online) Coercive field vs external field frequency at various temperatures, in a linear (a) and semilog plot (b). Every value is mediated over $1000$ realizations corresponding to different disorder configurations
The solid lines on the $T=0$ curves are a linear fit.}
\label{Hcvsfreq_log}
\end{figure}

The first issue to be addressed in dynamic hysteresis is clearly the effect 
of the external field frequency on the hysteresis. In Fig.~\ref{T=0}, 
we show hysteresis loops obtained at $T=0$ for various frequencies.
As expected from experiments and general considerations, 
small (high) frequencies correspond to narrow (large) cycles.
To quantify this observation we can focus on the coercive field behavior. 
In Fig. \ref{Hcvsfreq_log} we show the dependence of $H_c$ on the field frequency $dH/dt=\dot{H}$.

\begin{figure}[h]
\centerline{\psfig{figure=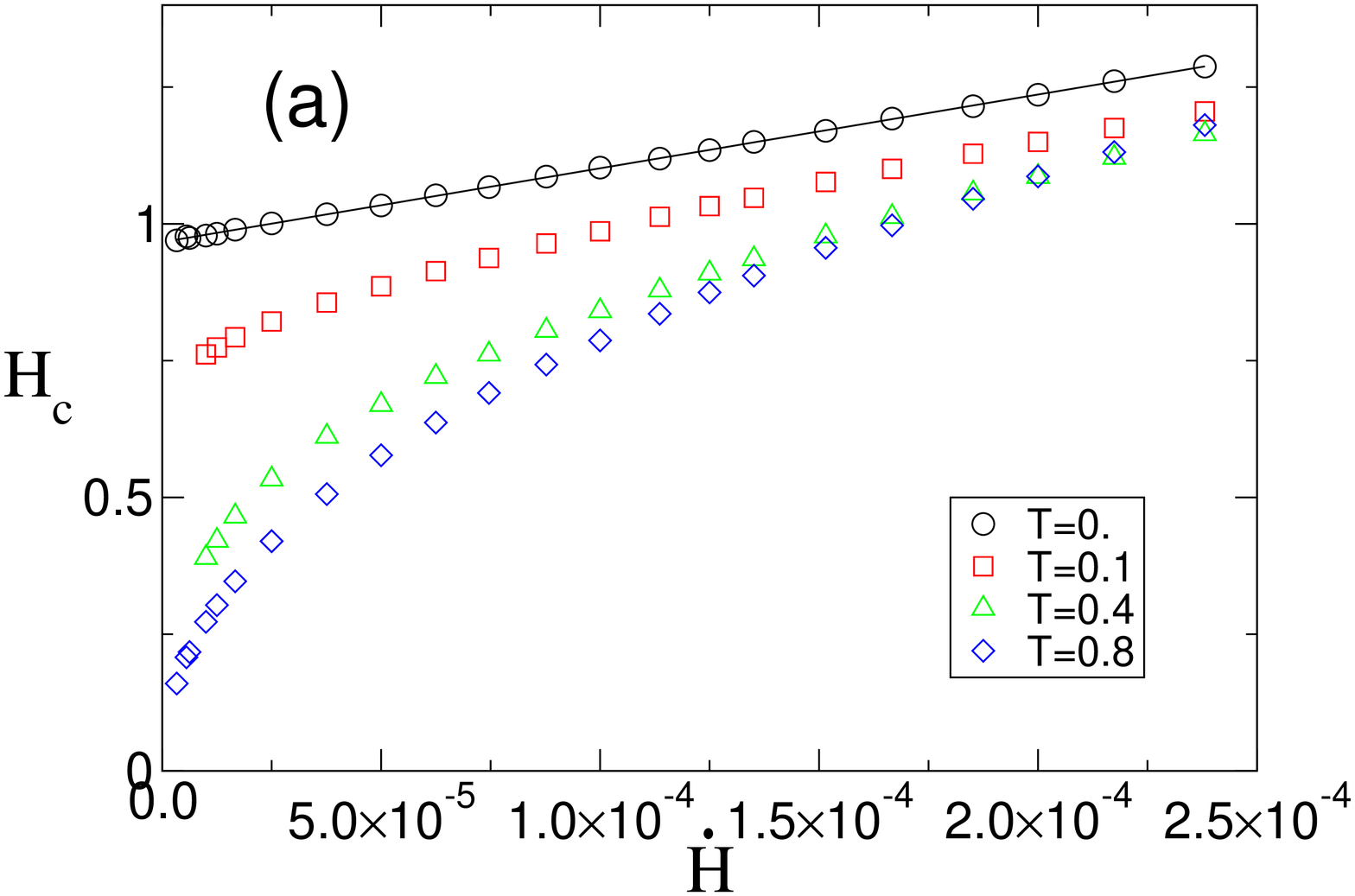,width=8cm,angle=-90,clip=}}
\label{Hcvsfreq}
\end{figure}

At $T=0$ (for a discussion of the behavior at non vanishing temperatures, see the following subsection),
$H_c$ shows an increasing linear dependence on $\dot{H}$, of the form
$H_c=H_p+A\dot{H}$, where $A$ is a constant. This means that in the adiabatic limit (low frequencies)
$H_c$ goes to a non-vanishing value $H_p$ (as clearly shown in
the semilogarithmic plot of Fig.\ref{Hcvsfreq_log}b), that we can interpret 
as the pinning dominated  quasistatic component due to structural disorder and anisotropy,
while the linear behavior of $H_c$ in the high frequencies regime 
represents the domain wall dominated dynamic contribution.

\begin{figure}[h]
\centerline{\psfig{figure=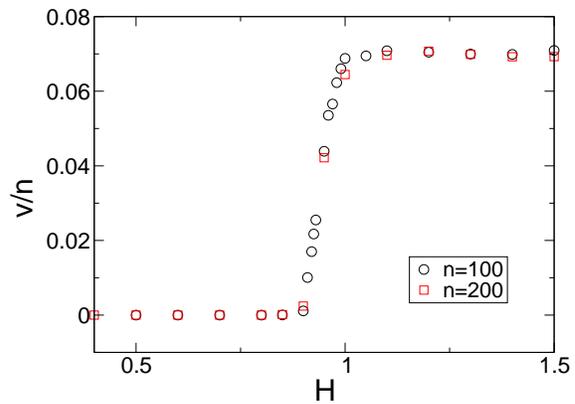,width=8.3cm,angle=-90,clip=}}
\caption{(Color online) Velocity of the domain wall, normalized to
the number of unit segments of the {\it zigzag}, as a function of an
external constant field. Each point is an average over 2000 different
disorder configurations.}
\label{velocita}
\end{figure}

This result is a particular case ($\alpha=1$) of the law
\begin{equation}
H_c=H_p+A\dot{H}^\alpha,
\label{eq:alpha}
\end{equation}
suggested on an experimental basis in Ref.~\onlinecite{nistor} (but
with a different exponent $\alpha$) and by the theory presented in
Ref.~\onlinecite{lyuksyutov}.  In the model of
Ref.~\onlinecite{lyuksyutov}, the exponent $\alpha$ is related to the
scaling exponent $\beta$ associated with the depinning transition of
the domain wall. In particular, it is assumed that under a constant
applied field the domain wall velocity $v \propto dM/dt$ follows
\begin{equation}
v = C(H-H_p)^\beta,
\label{eq:natter}
\end{equation}
for $H$ slightly larger than $H_p$, while $v$ vanishes for $H<H_p$.
Using Eq.~(\ref{eq:natter}) as a constitutive law, one can readily
show that the dynamic coercive field scales as in Eq.~(\ref{eq:alpha})
with $\alpha=1/(\beta+1)$. The limit $\alpha=1$ corresponds to
$\beta\to 0$ or else to a gap in the domain wall velocity as a
function of the external magnetic field. A very sharp dependence of
the velocity on the field is indeed observed in our model
(see Fig. \ref{velocita}) around the coercive field, where the segments of
the {\it zigzag} wall begin to move.  This is probably a strong
pinning effect: due to the {\it zigzag} structure of the wall 
the system is trapped by strong anisotropy barriers and collective
effects, typically leading to a continuous depinning, are 
suppressed.

\subsubsection{Temperature dependence}

\begin{figure}[h]
\centerline{\psfig{figure=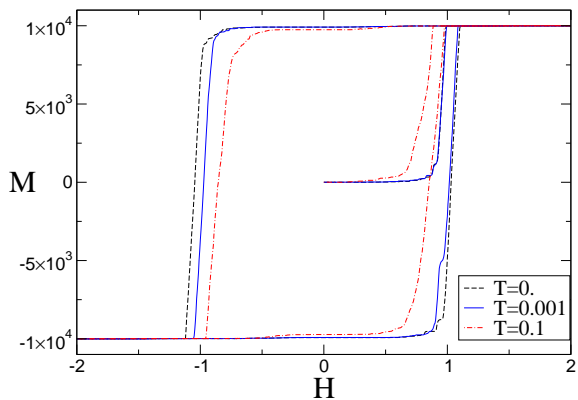,width=8.5cm,angle=-90,clip=}}
\caption{(Color online) Some hysteresis cycles for various temperature. The external fixed field time rate is $\Delta t=100$.}
\label{tempvar}
\end{figure}

\begin{figure}[h]
\centerline{\psfig{figure=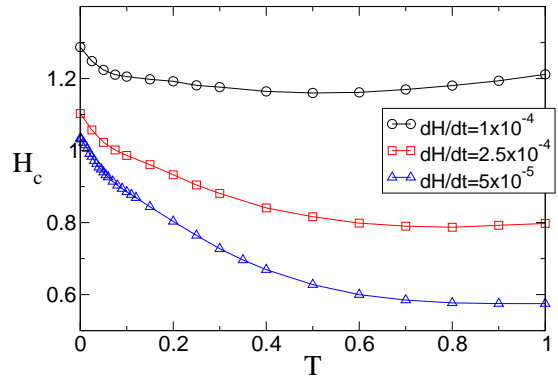,width=8cm,angle=-90,clip=}}
\caption{(Color online) Coercive field vs temperature for various external field time rates. Every value is mediated over $1000$ realizations corresponding to different disorder configurations.}
\label{HcvsT}
\end{figure}

Another interesting issue to analyze is the effect of temperature $T$
on dynamic hysteresis.  Even if thermal effects do not seem to be
relevant for most ferromagnetic thin films where $k_BT<\!\!<\Delta E$,
as it is easily checked, e.g., comparing $k_BT$ at room temperature,
where the experiments are typically performed, with an estimate of
$\Delta E \simeq E_{an}$ which could be obtained with the parameters
of $Fe/GaAs(001)$ given in Sec. \ref{comparison}. Thermal activation
could play a role at very low frequencies or in ultrathin
ferromagnetic films. In general terms, the increase of the temperature
(at reasonably low frequency, see Fig. \ref{HcvsT}) acts on the
hysteresis cycles shape in an similar way as the decrease of the
frequency (see Fig. \ref{tempvar}).  Since a temperature increase
enhances the probability for the wall to overcome energy barriers, at
high (low) temperature hysteresis loops will be large
(narrow). However at very high frequencies, the system is not able to
readily respond to the external field, and the decreasing dependence
of $H_c$ on increasing temperatures is violated, as shown in
fig. \ref{HcvsT}.  This explains the crossover between the curves with
$T=0.4$ and $T=0.8$ (Fig. \ref{Hcvsfreq_log}).  

In Fig. \ref{Hcvsfreq_log} we show the dependence of $H_c$ vs $\dot{H}$
at various non-zero temperatures.  It is interesting to note that our
simulations predict that the high $T$ behavior of $H_c$ vs $\dot{H}$
deviates from the linear behavior established at $T=0$.
This can be understood from  general considerations, 
since when $T>0$ thermal activation will lead creep domain wall motion
even for $H<H_p$. Hence, according to the theoretical analysis 
presented in Ref.~\onlinecite{nattermann2001} at low frequencies 
the dynamic coercive field will deviate from Eq.~(\ref{eq:alpha}) and decay
as $H_c \sim 1/[\log (\dot{H})]^{1/\mu}$ where $\mu$ is a creep
exponent. This result is consistent with  Fig. \ref{Hcvsfreq_log}b,
although the limited scaling range does not allow for a quantitative
confirmation.

\subsubsection{Thickness dependence}

Finally, we address to the film thickness role in dynamic hysteresis. 
As it can be seen in Eq.~\ref{eps_riscalati}, for sufficiently large $\epsilon$ 
above the purely disorder dominated regime discussed in section \ref{epsi}, the energy barrier increases 
linearly with $\epsilon$ at $T=0$, and the coercive field  
does so as well. 

\begin{figure}[h]
\centerline{\psfig{figure=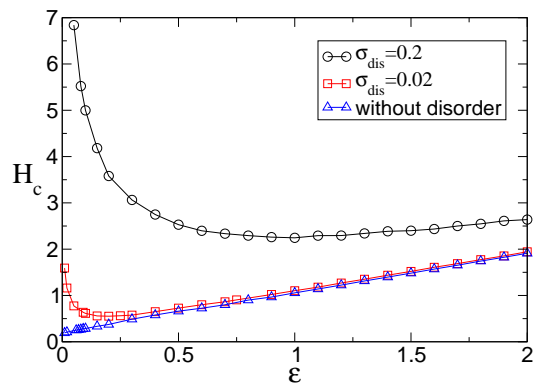,width=8cm,angle=-90,clip=}}
\caption{(Color online) Coercive field vs film thickness for various values of the $\sigma$ of the disorder Gaussian distribution. Every value is mediated over $200$ realizations corresponding to different disorder configurations.}
\label{Hcvseps}
\end{figure}

\begin{figure}[h]
\centerline{\psfig{figure=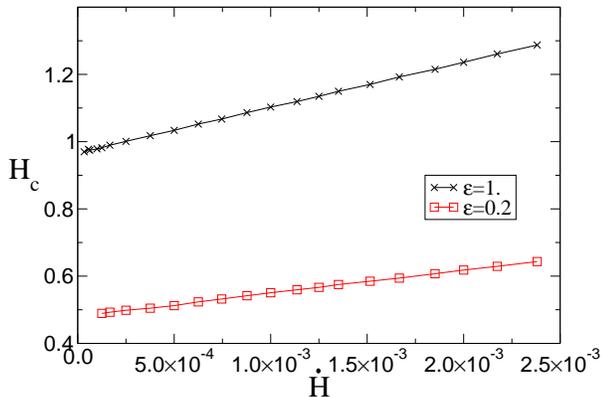,width=8.3cm,angle=-90,clip=}}
\caption{(Color online) Coercive field vs frequency $\dot{H}$ at $T=0$ for two different thicknesses. 
The linear behavior is not affected by the thickness. Compare the
present figure with the experimental results reported in
Fig.~\protect\ref{sperimentale_fit}}
\label{2eps}
\end{figure}

Turning to a more quantitatively discussion, at $T=0$ the results of
Eq.~\ref{eps_riscalati} are confirmed by the simulations 
for various $\epsilon$ 
summarized in Fig. \ref{Hcvseps}. In the two upper curves, 
where the disorder energy term is not negligible, 
we notice three regimes: a divergence proportional to $1/\sqrt{\epsilon}$
at very low $\epsilon$, due to the disorder term 
and linear regime  for high $\epsilon$ values, due to the dipolar term
(the regime independent from $\epsilon$ that would be due to the
anisotropy energy and the external field, cannot be seen clearly in that figure). 
Moreover, from the lower curve of Fig. \ref{Hcvseps} 
(the one without disorder) we confirm
that the low-$\epsilon$ divergence is due to the disorder term.
Otherwise, the thickness dependence of the coercive field does not affect 
the linearity of the frequency dependence of $H_c$, which remains valid 
for every $\epsilon$ value, as is seen in Fig. \ref{2eps} (at $T=0$).

\section{comparison with experiments}

\begin{figure}[h]
\centerline{\psfig{figure=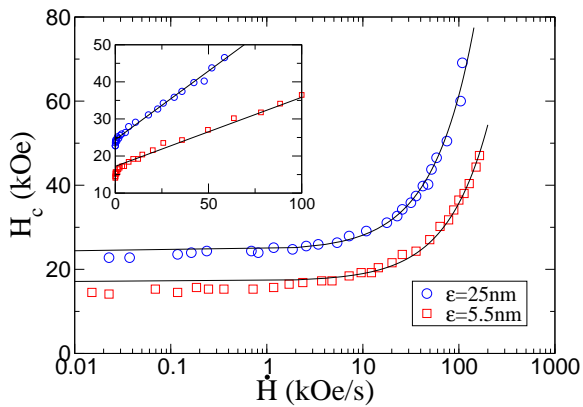,width=8.3cm,angle=-90,clip=}}
\caption{(Color online) Coercive field as a function of the external field frequency from
experiments on Fe/GaAs thin films for two different thicknesses. 
The data is fitted properly by linear functions (solid lines), 
but when plotted
in semi-logarithmic scale they may suggest a crossover at $\dot{H} \simeq 20$. 
In the inset: a linear plot of the frequencies up to $100 kOe/s$.  
The deviation from linear behavior is quite small and could be do to
thermal activation. Data are obtained from Ref.~\protect\onlinecite{bland1}.}
\label{sperimentale_fit}
\end{figure}

The linear dependence of $H_c$ vs $\dot{H}$ is consistent with the experiments
\cite{bland1}, at least in the high frequency regime, as can be seen from 
Fig. \ref{sperimentale_fit}, even though there the authors 
try to identify two dynamical regimes in a semi-logarithmic plot, 
attributed to two different dynamic processes, 
namely domain nucleation and domain wall propagation.
This is probably an artifact of the logarithmic scale. 
Our conclusion is rather different:
as it is quite clear from figure \ref{sperimentale_fit},
the experimental results \cite{bland1} on the frequency dependence 
of the coercive field 
could be properly 
explained by a model based on a single propagating domain wall, 
which is the only 
dynamic process that we have considered, without invoking domain nucleation.

In the low frequency regime, the experimental data show a deviation
from the linear dependence.  Even if this deviation is quite small and
could be due simply to the experimental error, our results suggest
that it could be interpreted by means of thermal activated motion. This
interpretation seems to be supported by the fact that the deviation
is stronger for the thinner sample, as it would be expected if thermal
effects where the cause. We notice as well that the data reported in 
Ref.~\cite{moore2} for even thinner films of Fe/GaAs show a logarithmic
decrease of $H_c$ at low frequency, not observed for thicker samples
where $H_c$ tends instead to a  constant as in Fig.~\ref{sperimentale_fit}.
Finally our model accounts as well for the fact that 
the value of $H_c$ for the thinner sample 
(in our notation, the one with $\epsilon=55$\AA) is
smaller than the value measured for the thicker one
($\epsilon=250$\AA) at the same field frequency $\dot{H}$.  
As a word of caution it is important to remark here that
the only sample for which a {\it zigzag} domain structure is clearly
reported is the one with $250$\AA$\;\!$ 
thickness.

\section{conclusions}
\label{conclusions}

The dynamics of two dimensional ferromagnetic systems is still under
debate, both on the theoretical and the experimental side. A crucial
issue is the description of the dynamic hysteresis, that is related to
power losses and thus plays an important role in several technological
applications. Here, we have analyzed ferromagnetic thin films with
{\it zigzag} domain walls, arising when the magnetization vectors in
two nearest-neighbor domains meet head-on at the wall that separates
them.  To investigate dynamic hysteresis, we have studied the motion
of {\it zigzag} domain walls by developing a simple discrete model
based on the interplay between dipolar and anisotropy energy
contributions, in presence of structural disorder.  Under some simple
approximations one can estimate some experimentally relevant
quantities, such as the typical {\it zigzag} half-period and the
coercive field, which turn out to be in quantitative agreement with
experimental observations.

Although quite simplified, our model allows to recover the behavior of
coercive field $H_c$ in dynamic hysteresis. We have studied the
dependence of $H_c$ on the applied magnetic field frequency $\dot{H}$
at $T=0$ and found that the coercive field scales as $H_c=H_p +
A\dot{H}$.  This linear behavior is in good agreement with experiments
\cite{bland1}, which we can thus explain by means of pure domain wall
propagation model, without the need to invoke other dynamic processes
as domain nucleation. We have also simulated hysteresis at $T>0$, even
if this case is probably not relevant for most ferromagnetic thin
films, where thermal effects are negligible.  We show that high
temperature at low frequency induces narrow loops and the coercive
field decreases with respect to the $T=0$ case, while at high
frequency the situation is less intuitive due to the delay between the
system response and the external driving field.  We have also studied
the dependence of the coercitivity from the film thickness $\epsilon$.
The behavior indicated by the simulations is explained by simple
analytical considerations.  For small disorder, we find that the
thickness does not affect the rate dependence of $H_c$ at $T=0$. It is
interesting to remark that our model could be applicable to
ferroelectric materials which are known to show as well {\it zigzag}
domain walls. It would be very interesting to compare our results with
experiments in this case as well.

\section*{Acknowledgments}
We would like to thank G. Durin and A. Magni for useful discussions.

\clearpage

\section{appendix}

Analitical calculation of magnetostatic energy: the
result of equation (\ref{Eij}) is given by

\begin{eqnarray}
\displaystyle
E_{ij} &=& 8M_s^2\epsilon^2\times 
\nonumber\\
\displaystyle
&&\left\{\delta(m_i,m_j)\left[g_1(m_i,q_i-q_j,m_j,jp,ip)\right.\right.
\nonumber\\
&&-g_1(m_i,q_i-q_j,m_j,jp,(i+1)p)
\nonumber\\
&&-g_1(m_i,q_i-q_j,m_j,(j+1)p,ip)
\nonumber\\
&&\left.+g_1(m_i,q_i-q_j,m_j,(j+1)p,(i+1)p)\right]
\nonumber\\
&&+\delta(m_i,-m_j)\left[g_2(m_i,q_i-q_j,m_j,jp,ip)\right.
\nonumber\\
&&-g_2(m_i,q_i-q_j,m_j,jp,(i+1)p)
\nonumber\\
&&-g_2(m_i,q_i-q_j,m_j,(j+1)p,ip)
\nonumber\\
&&\left.\left.+g_2(m_i,q_i-q_j,m_j,(j+1)p,(i+1)p)\right]\right\},
\nonumber\\
\label{risultato}
\end{eqnarray}

where

$$
\left\{
\begin{array}{lcr}
\displaystyle
g_1(m,q-q',m,x,x')=\frac{1}{1+m^2}\times
\\\\
\displaystyle
\,\,\,\,\,\,\,\,\,\,\,\,\left\{|{\bf r}-{\bf r}'|-\frac{a({\bf r}-{\bf r}')}{2}\ln\left[\frac{|{\bf r}-{\bf r}'|-a({\bf r}-{\bf r}')}{|{\bf r}-{\bf r}'|+a({\bf r}-{\bf r}')}\right]\right\}
\\\\\,\,\,\,\,\,\,\,\,\,\,\,\,\,\,\,\,\,\,\,\,\,\,\,\,\,\,\,\,\,\,\,\,\,\,\,\,\,\,\,\,\,\,\,\,\,\,\,\,\,\,\,\,\,\,\,\,\,\,\,\,\,\,\,\,\,\,\,\,\,\,\,\,\,\,\,\,\,\,\,\,\,\,\,\,\,\,\,\,\,\,\,\,\,\,\,\,\,\,\,\,\,\,\,\,\,\,\,\forall\,\,\,\,\,\,\,\,\,\,\,\, q\neq q'
\\\\
\displaystyle
g_1(m,q-q',m,x,x')=\frac{(-x+(x-x')\ln(x-x'))}{\sqrt{1+m^2}}
\\\\\,\,\,\,\,\,\,\,\,\,\,\,\,\,\,\,\,\,\,\,\,\,\,\,\,\,\,\,\,\,\,\,\,\,\,\,\,\,\,\,\,\,\,\,\,\,\,\,\,\,\,\,\,\,\,\,\,\,\,\,\,\,\,\,\,\,\,\,\,\,\,\,\,\,\,\,\,\,\,\,\,\,\,\,\,\,\,\,\,\,\,\,\,\,\,\,\,\,\,\,\,\,\,\,\,\,\,\,\mbox{if}\,\,\,\,\,\,\,\,\,\,\,\, q=q'
\end{array}
\right.
$$

$$
g_2(m,q-q',m,x,x')=b({\bf r}-{\bf r}')\ln\left[\frac{|{\bf r}-{\bf r}'|+a({\bf r}-{\bf r}')}{|{\bf r}-{\bf r}'|-a({\bf r}-{\bf r}')}\right]
$$
$$
+b({\bf r}'-{\bf r})\ln\left[\frac{|{\bf r}-{\bf r}'|+c({\bf r}-{\bf r}')}{|{\bf r}-{\bf r}'|-c({\bf r}-{\bf r}')}\right]
$$

and

$$
\left\{
\begin{array}{lc}
\displaystyle
a({\bf r}-{\bf r}')=\frac{x'-x+m^2(x'-\frac{m'}{m}x)-m'(q-q')}{\sqrt{1+m^2}}\\
\\
\displaystyle
b({\bf r}-{\bf r}')=\frac{q-q'+2mx}{4m\sqrt{1+m^2}}
\\
\\
\displaystyle
c({\bf r}-{\bf r}')=\frac{x-x'+m^2(x+x')+m(q-q')}{1+m^2}
\end{array}
\right.
$$

and $\delta(m,m')$ is Kronecker's delta.

\begin{figure}[h]
\centerline{\psfig{figure=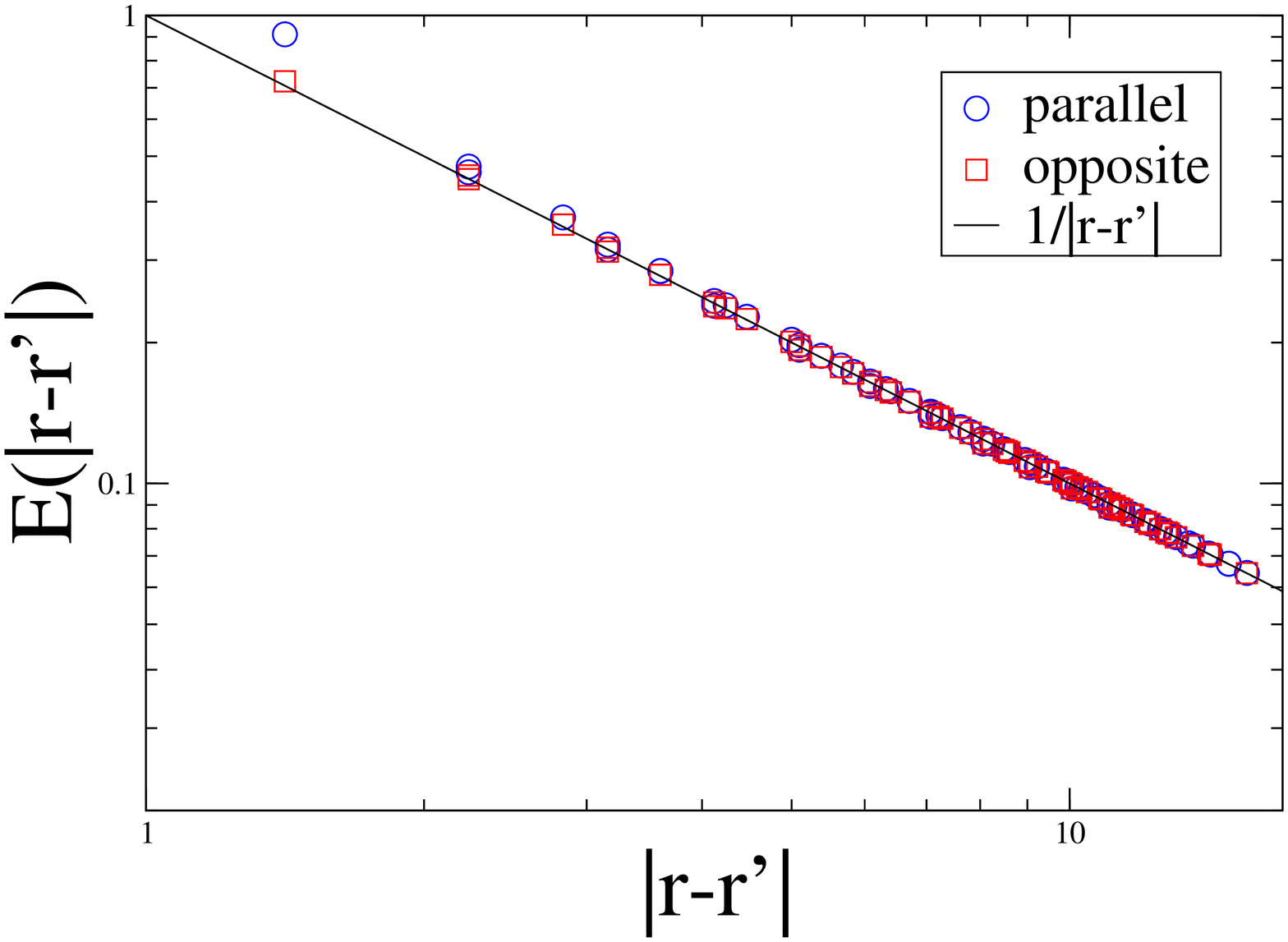,width=7.5cm,clip=,angle=-90}}
\caption{(Color online) Interaction magnetostatic energy between two generic segments with parallel or opposite slopes as a function of the distance between their centers of mass.}
\label{E_vs_r}
\end{figure}

In Fig. \ref{E_vs_r} we plot the function $E_{ij}$ 
(for unitary $8M_s^2\epsilon^2$)
 as a function of the distance between the centers of 
mass of the segments $i$ and $j$. As could be seen, 
the value of magnetostatic interaction
energy is mainly proportional to $1/|{\bf r}-{\bf r}'|=1/r_{ij}$ 
for each pair of $i$ and $j$ but the nearest-neighbour, where magnetic charges 
are sufficiently close to each other to experience the very shape of their 
spatial distribution.

\end{document}